\documentclass[superscriptaddress,twocolumn,floatfix,longbibliography]{revtex4-2}

\usepackage{graphicx}
\usepackage{dcolumn}
\usepackage{bm}
\usepackage{url}

\usepackage{amsfonts}
\usepackage{wrapfig}
\usepackage{bbm}

\def \beq {\begin{equation}}
	\def \eeq {\end{equation}}
\def \ba {\begin{eqnarray}}
	\def \ea {\end{eqnarray}}

\usepackage[latin1]{inputenc}
\usepackage{graphicx}
\usepackage{amsmath}
\usepackage{amsthm}
\usepackage{bm}
\usepackage{layout}
\usepackage{float}
\usepackage{amssymb}%
\usepackage{color}
\usepackage{soul}
\usepackage{mathtools}
\usepackage[colorlinks=true,citecolor=blue]{hyperref}

\theoremstyle{definition}
\newtheorem{theorem}{Theorem}

\newtheorem{lemma}{Lemma}

\usepackage[capitalise,compress]{cleveref}

\newcommand{\ketbrad}[1]{|#1\rangle\!\langle #1|}

\newcommand{\mean}[1]{\langle#1\rangle}

\newcommand{\inp}{\textrm{in}}
\newcommand{\out}{\textrm{out}}
\newcommand{\Uin}{\mathcal{U}_\inp}
\newcommand{\Uout}{\mathcal{U}_\out}
\newcommand{\bin}{b_\inp}
\newcommand{\bout}{b_\out}
\newcommand{\UP}{\mathcal{U}_{\rm P}^{\otimes n}}
\newcommand{\UC}{\mathcal{U}_{\rm C}}
\newcommand{\UPone}{\mathcal{U}_{\rm P}}
\newcommand{\Tr}{\textrm{Tr}}
\newcommand{\E}{\mathcal{E}}
\newcommand{\norm}[1]{\left\|#1\right\|}
\newcommand{\abs}[1]{\left | #1 \right|}
\newcommand{\rank}{\text{rank}}
\newcommand{\supp}{\text{supp}}
\def\ket#1{\left| #1\right>}
\def\bra#1{\left< #1\right|}

\usepackage{xcolor}

\begin{document}
	
\title{Shadow process tomography of quantum channels}
	
\author{Jonathan Kunjummen}
\email{jkunjumm@umd.edu}
\affiliation{Joint Center for Quantum Information and Computer Science, NIST/University of Maryland, College Park, Maryland 20742, USA}
\affiliation{Joint Quantum Institute, NIST/University of Maryland, College Park, Maryland 20742, USA}
	
\author{Minh C. Tran}
\affiliation{Center for Theoretical Physics, Massachusetts Institute of Technology, Cambridge, Massachusetts 02139, USA}
\affiliation{Department of Physics, Harvard University, Cambridge, Massachusetts 02138, USA}
\affiliation{Joint Center for Quantum Information and Computer Science, NIST/University of Maryland, College Park, Maryland 20742, USA}
\affiliation{Joint Quantum Institute, NIST/University of Maryland, College Park, Maryland 20742, USA}

\author{Daniel Carney}
\affiliation{Physics Division, Lawrence Berkeley National Laboratory, Berkeley, California 94720, USA}
\author{Jacob M. Taylor}
\affiliation{Joint Center for Quantum Information and Computer Science, NIST/University of Maryland, College Park, Maryland 20742, USA}
\affiliation{Joint Quantum Institute, NIST/University of Maryland, College Park, Maryland 20742, USA}

\date{\today}

	
\begin{abstract}
Quantum process tomography is a critical capability for building quantum computers, enabling quantum networks, and understanding quantum sensors. Like quantum state tomography, the process tomography of an arbitrary quantum channel requires a number of measurements that scales exponentially in the number of quantum bits affected. However, the recent field of shadow tomography, applied to quantum states, has demonstrated the ability to extract key information about a state with only polynomially many measurements. In this work, we apply the concepts of shadow state tomography to the challenge of characterizing quantum processes. We make use of the Choi isomorphism to directly apply rigorous bounds from shadow state tomography to shadow process tomography, and we find additional bounds on the number of measurements that are unique to process tomography. Our results, which include algorithms for implementing shadow process tomography, enable new techniques including evaluation of channel concatenation and the application of channels to shadows of quantum states. This provides a dramatic improvement for understanding large-scale quantum systems.
\end{abstract}
\maketitle

\section{Introduction}

Characterizing dynamical processes in quantum mechanics is a ubiquitous task. In the most general setting, a quantum process is described by a quantum channel--a linear map which takes a quantum state and maps it to another. In quantum computing, the execution of a quantum circuit is an attempt to faithfully produce a good approximation to a desired quantum channel, which represents the integrated effect of the specific experiment on the underlying quantum state. In quantum networking, quantum processes act during information transmission \cite{gisin2007quantum}, characterizing uniquely quantum phenomena like the generation of entanglement while also leading to fundamental limitations to noiseless communication \cite{holevo1973bounds}. Beyond technological implications, understanding the properties of quantum channels underlies fundamental questions in disparate fields, including 
the characterization of entanglement properties of gravity \cite{2013arXiv1311.4558K,2017PhRvL.119x0401B,2017PhRvL.119x0402M,Carney:2018ofe,2021PRXQ....2c0330C}.

Quantum channels can, in principle, dramatically change arbitrary quantum states, having an input and an output of a density matrix. Thus they require $O(d^4)$ different parameters to describe their action on a $d$-dimensional Hilbert space. Fully characterizing an arbitrary channel requires a number of measurements that is polynomial in the Hilbert space dimension, e.g., exponential in the number of qubits, even when channels are restricted to a smaller set such as unitary or nearly unitary channels~\cite{gross_quantum_2010}. Quantum process tomography is the detailed, and practical, exploration of how to do this as effectively as possible \cite{chuang1997prescription,branderhorst2009simplified,shabani2011efficient,flammia_quantum_2012}. Further difficulties in channel reconstruction also arise from the requirements that their action preserve the trace and positivity of density matrices~\cite{Huang:2021ydj}.

Of course, the challenges that occur for quantum channels also occur for quantum states. In the past several years, there have been advances in reducing the number and type of measurements to better estimate properties of interest of quantum states by leveraging a concept from classical tomography: creating projective inverses of quasi-probability distributions. This extension to quantum states, denoted shadow quantum state tomography \cite{aaronson_shadow_2018,huang_predicting_2020}, has already shown a dramatic reduction in the number of measurements necessary to efficiently estimate key properties, such as purity and expectation values of local operators \cite{cian_cross-platform_2021,struchalin_experimental_2021,zhang_experimental_2021}. Furthermore, these techniques use simple, easy to parallelize means of data storage and processing, and maximally leverage all experimental information per copy of the state estimated~\cite{2021arXiv210701060S}. 

Here we apply the concept of shadow quantum state tomography (which we call shadow state tomography, dropping the `quantum') to quantum channels. We show that many of the key results for shadow state tomography can be directly imported to quantum process tomography using `process shadows'. Our approach leverages the Choi isomorphism \cite{jamiolkowski_linear_1972,choi_completely_1975,leung2003choi}, which has the natural interpretation of representing the action of a channel on one half of a maximally entangled state. However, key differences between inputs and outputs, as well as restrictions on Choi states, lead to a need to develop new measurement bounds, which we provide with a key theorem. 
Furthermore, we also show that shadows of quantum channels have additional useful properties, including their application to shadows of states, and the ability to concatenate multiple shadow channels to create a new, quasi-shadow channel describing the more complex outcome. 

Our paper begins with a brief introduction of state tomography and shadow state tomography, before leveraging the Choi isomorphism to provide an algorithm for shadow process tomography. We then prove the key theorem of this work describing rigorous bounds on measurement needs for accurately estimating the properties of a channel. Focusing on the case of preparing individual qubit Pauli eigenstates and measuring individual qubit outcomes, we show how to practically implement the shadow process tomography algorithm and consider the effects of applying a shadow process outcome to a shadow state. We find that both this case, and the case of concatenating channels together, leads to quasi-shadows with negative probabilities, and conjecture that an opportunity exists to reduce these quasi-shadows to maintain efficient performance. We conclude with numerical examples of our protocols applied to several different types of few qubit channels.


\section{Traditional and shadow tomography}

We start with a state $\rho$ taken to be a density matrix for $n$ qubits. Choosing a basis for operators $\{ O_i \}$, we can write $\rho = \sum_i q_i O_i + \mathbb{I}/2^n$. Given this representation, we consider a complete set of traceless measurements $\{ M_i \}$. Informational completeness tells us that we can find the $q_i$'s given expectation values of the $M_i$'s under $\rho$:
\begin{equation}
	\mean{M_i}_{\rho} = \sum_j q_j {\rm Tr}[M_i O_j]
\end{equation}
If the matrix $\mathcal{M}_{ij} = {\rm Tr}[M_i O_j]$ is invertible, then the linear system of equations defined above can be solved. In what follows, we will take the $O$'s to be equal to the $M$'s, and let them be relatively simple to implement, e.g., all Pauli strings over $n$ qubits for a $d = 2^n$ dimensional space (the `Clifford set'), or all single-qubit Paulis operators (the `Pauli set'). In practice, the finite errors in expectation values due to sampling only a finite number of times leads to errors in the associated estimate of the density matrix, and nonlinear reconstruction techniques need to be employed to create physical (that is, completely positive and unit trace) density matrices that are also consistent with the measurement outcomes. 

Here, we are instead focused on the question of how we can best use the measurement results, and use the language of shadow tomography~\cite{aaronson_shadow_2018,huang_predicting_2020} to explore this question.
In shadow tomography, we consider a scenario where finding $\mathcal{M}^{-1}$ is straightforward, and furthermore, where the sum can be done implicitly.  Take a basis for Hilbert space $\{ \ket{b} \}$ where $b$ are $n$ bit strings representing eigenstates of individual qubits in the $Z$ basis. Let $\{ U \}$ be a set of unitary operators and associating with each $U$ there is a probability $p_U\geq 0$. We denote this ensemble of unitaries by $\mathcal U$. We can define
\begin{equation}
	\mathcal{M}_{\mathcal U}(\rho) = \mathbb E_U \sum_b {\rm Tr}[U^\dag \ketbrad{b} U \rho] \  U^\dag \ketbrad{b} U.
\end{equation}
We shall drop the subscript $\mathcal U$ if there is no ambiguity. 
We see that $\mathcal{M}$ is a linear operator. When $\mathcal{M}^{-1}$ exists, then the above equation has a simple interpretation. Let $p_{U,b} =p_U {\rm Tr}[U^\dag \ketbrad{b} U \rho]$. Then
\begin{equation}
	\rho = \sum_{U,b} p_{U,b} \mathcal{M}^{-1}[U^\dag \ketbrad{b} U ]
\end{equation}
up to normalization. That is, we take advantage of the linearity to do the inverse inside the sum. We see that the $p_{U,b}$ are positive numbers and can be interpreted as a probability. In general, we only sample over the sum on $U$ and $b$, and thus the finite sample version of $\rho$ only is expected to reproduce $\rho$ in expectation.



For the case of single qubits and Pauli measurements, this inverse is straightforward to write down:
\begin{equation}
	\mathcal{M}^{-1}_{\UPone}(A) = 3 A - \Tr(A) \mathbb I.\label{eq:inverse-pauli}
\end{equation}
for any operator $A$.
Here, we use $\UPone$ to denote the ensemble of uniformly random single-qubit Pauli operators. 
The inverse map for $n$-qubit Pauli measurements is simply $\mathcal M_{\UP}^{-1} = (\mathcal M_{\UPone}^{-1})^{\otimes n}$.
Similarly, for $U$ taken from an ensemble $\UC$ of arbitrary $n$-qubit Clifford circuits, 
\begin{equation}
	\mathcal{M}^{-1}_{\UC}(A) = (2^{n}+1) A - \Tr(A)\mathbb{I}.
	\label{eq:inverse-clifford}
\end{equation}

In practice, we take $m$ copies of $\rho$, and for each copy, choose a random $U$ (corresponding to setting a measurement basis) from our set. We then measure in the logical basis to get an outcome $b$. Let $Q$ be the set of $m$ such pairs of $U$ and $b$ and $ \hat \sigma_j \equiv \mathcal M^{-1}(U_j^\dag \ketbrad{b_j} U_j)$ for $j = 1,\dots,m$. Then 
\begin{equation}
	\tilde{\sigma}_m = \frac{1}{m} \sum_{j = 1}^m \hat \sigma_j
\end{equation}
is a finite-sample estimator of $\rho$.
In the limit of large sample size, $\lim_{m\rightarrow \infty} \tilde \sigma_m = \rho$.

Critically, work on this topic~\cite{huang_predicting_2020} has shown that only polynomially (in number of qubits) many such measurements are required to reproduce expectation values of certain observables.
Given an observable $O$, let $\hat o_j = \Tr(\hat \sigma_j O)$ for $i = 1,\dots,m$.
We then compute the median of means $\tilde o(N,K)$ by dividing $Q$ into $K$ subsets, each of size $N = m/K$, and calculating the median of the means of $\hat o_j$ in the subsets. The results of these are encompassed in the theorem from the above work, which we restate here:

\begin{theorem}[State shadow tomography~\cite{huang_predicting_2020}]\label{thm:state-tomo}
    Given a set of density matrices $\Omega$, a density matrix $\rho\in \Omega$, a unitary ensemble $\mathcal U$, a collection of $M$ observables $O_1,\dots,O_M$, and any $\epsilon,\delta\in[0,1]$,
    \begin{align}
		&K = 2\log(2M/\delta),\\
		&N = \frac{34}{\epsilon^2}\max_{1\leq \mu \leq M} 
		\min\bigg\{\norm{O_\mu - \frac{\Tr(O_\mu)}{2^n}\mathbb I}^2_{\text{shadow},\mathcal U,\Omega},\nonumber\\
		&\qquad\qquad\qquad\qquad\qquad\qquad   \norm{O_\mu}_{\text{shadow},\mathcal U,\Omega}^2\bigg\},
    \end{align}
    a set of $m = NK$ independent shadows are sufficient for
    \begin{align}
		\abs{\tilde o_\mu(N,K) - \Tr(O_\mu \rho)}\leq \epsilon \quad \forall 1\leq \mu\leq M
    \end{align}
    with probability at least $1-\delta$.
    Here, the shadow norm of an observable $O$ with respect to an unitary ensemble $\mathcal U$ and a set of density matrices is defined as
    \begin{align}
		&\norm{O}_{\text{shadow},\mathcal U,\Omega} ^2 
		\coloneqq \max_{\sigma\in \Omega} \sum_{b\in\{0,1\}^n} \bra{b}U\sigma U^\dag \ket{b} \nonumber\\
        &\qquad\qquad\qquad\qquad\qquad\times\bra{b} U \mathcal M^{-1}_{\mathcal U} (O) U^\dag \ket{b}^2,
    \end{align}
    where the maximization is over all density matrices $\sigma$ in $\Omega$.
    We will drop the subscript $\Omega$ if $\Omega$ is the set of all density matrices of $n$ qubits.
\end{theorem}
In addition, Ref.~\cite{huang_predicting_2020} provides useful bounds on the shadow norms when $\mathcal U$ is either $\UC$ or $\UP$:
\begin{align}
    &\norm{O}_{\text{shadow},\UP}^2 \leq 4^{\supp(O)} \norm{O}^2,\label{eq:shadow-norm-pauli}\\
    &\norm{O-\frac{\Tr(O)}{2^n}\mathbb I}_{\text{shadow},\UC}^2 \leq 3 \Tr(O^2) \label{eq:shadow-norm-clifford}.
\end{align}
\Cref{eq:shadow-norm-pauli} proves that the shadow tomography is efficient in estimating expectation values of local (few-qubit) observables when $\mathcal U$ is an ensemble of random local Pauli operators. 
On the other hand, if we choose $U$ to be arbitrary Clifford circuits, \cref{eq:shadow-norm-clifford} shows that the sample complexity scales with the rank of the observable and will be small for calculating, for example, the fidelity to a pure reference state. Therefore, depending on our purpose, we would choose a different unitary ensemble.

\section{Process shadows}
Just as state tomography builds an estimate of a density matrix, process tomography builds an estimate of a quantum channel. Recall that a channel $\E(\rho)$ can always be written using a Kraus representation as
\begin{equation}
    \E(\rho) = \sum_{ij} \chi_{ij} K_i \rho K_j^\dag
\end{equation}
where $\{ K_i \}$ is a complete and orthogonal (under the trace inner product) set of operators for the vector space of matrices of the same dimension $d \times d$ as $\rho$, and $\chi^\dag = \chi$. The $\{K_i\}$ are often taken to be Pauli strings, though in principle other choices can be made. There is an additional constraint $\sum_{ij} \chi_{ij} K_j^\dag K_i = \mathbb{I}$.
	
Given this representation, in process tomography we can consider a series of experiments in which we prepare the density matrix in some (known) state drawn from $\{ \rho_i \}$ and find the expectation value of some set of observables $\{ M_i \}$. Formally,
\begin{equation}
    m_{ij} = {\rm Tr}[M_i \E(\rho_j)] = \sum_{kl} \chi_{kl} {\rm Tr}[M_i K_k \rho_j K_l^\dag]
\end{equation}
If we think of the vector space of matrices by combining $i,j$ in a label $a$ and $k,l$ into a label $b$ we see that if the matrix $\mathcal{K}_{ab} = {\rm Tr}[M_i K_k \rho_j K_l^\dag]$ is invertible then we can recover $\chi_{kl}$, just as we did for the density matrix in the previous section.
	
\subsection{Choi isomorphism}
We now ask: can the principles of shadow tomography be applied to process tomography?  In order to apply the shadow tomography framework to a process $\E$, we make our lives conceptually easier through the use of (the Choi version of) the Choi-Jamio{\l}kowski isomorphism to map a process into a density matrix. Specifically, we specialize to the case of finite dimension, where $\E: \mathcal{C}^{d \times d} \rightarrow \mathcal{C}^{d \times d}$ and is completely positive and trace-preserving. We define the unnormalized, maximally entangled state on a tensor product of two Hilbert spaces $A$ and $B$ each of dimension $d$ 
\begin{equation}
    \ket{\omega} = \sum_{n=0}^{d-1} \ket{n}_A \otimes \ket{n}_B 
\end{equation}
(We remark that this state can be prepared by starting in the logical $\ket{0}_A \otimes \ket{0}_B$ state, applying Hadamard to each qubit in the first register, then applying pair-wise CNOTs from the $i$th qubit of the first register to the $i$th qubit of the second register.) We then define the Choi state 
\begin{equation}\label{eq:Choi-state}
    \eta = (\mathcal{I}_A \otimes \E_B)[\ket{\omega} \bra{\omega}].
\end{equation}
We note that $\eta$ is a density matrix (completely positive and of finite trace) up to a normalization factor $d$.
	
\begin{figure}
    \centering
    \includegraphics[width=0.45\textwidth]{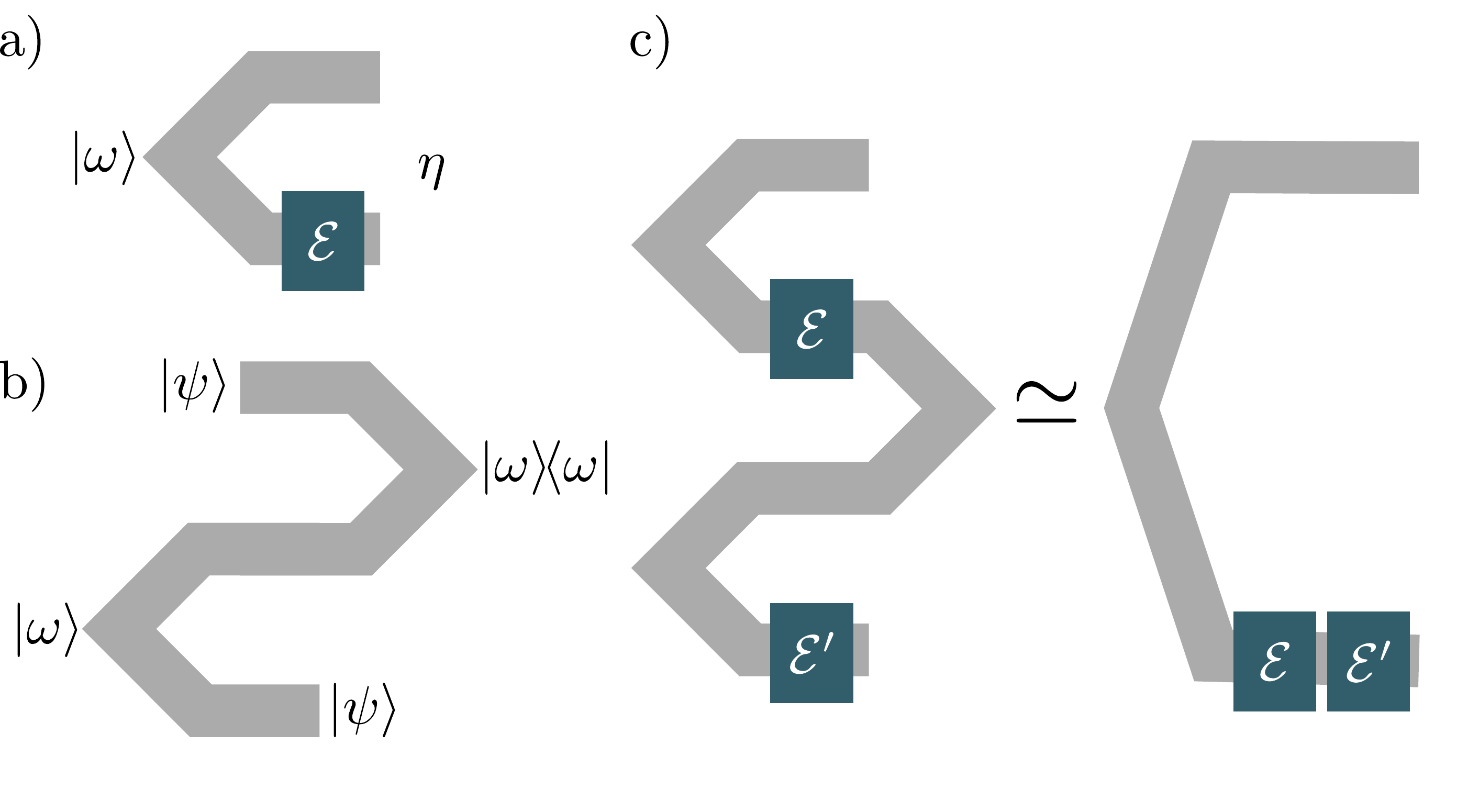}
    \caption{(a) The Choi isomorphism in circuit form, (b) its use for state teleportation, and (c) its application for channel concatenation. An opening pair of thick grey lines represents creation of the maximally-entangled state $\ket{\omega}$. A closing pair indicates (successful) projection of the state onto $\ket{\omega}$.}
    \label{f:choiiso}
\end{figure}

Given a Choi state $\eta$ of dimension $d^2 \times d^2$ and an input density matrix $\rho$ of size $d \times d$, we can find the result of $\E(\rho)$ as
\begin{equation}\label{eq:Choi-to-channel}
    \E(\rho) = {\rm Tr_A}[(\rho^T\otimes \mathbb I_B)\, \eta ]
\end{equation}
where we have taken the transpose of $\rho$ with respect to the logical basis (that is, the basis in which we defined $\ket{\omega}$). This has the (simple) interpretation that given a Choi state, we can teleport $\rho$ through $\E$ to find its result, as shown in \cref{f:choiiso}.
	
To understand the teleportation interpretation, consider an arbitrary state $\ket{\psi}_C = \sum_n c_n \ket{n}$. Starting with $\ket{\omega}_{AB} \ket{\psi}_C$ and projecting onto $\ket{\omega}_{BC}$ yields:
\begin{align}
    &\bra{\omega}_{BC} \ket{\omega}_{AB} \ket{\psi}_C \nonumber\\
    &=     \sum_{m,m',n} c_n \bra{m}_B \bra{m}_C \ket{m'}_A \ket{m'}_B \ket{n}_C \\
    &=   \sum_{m,n} c_n \bra{m}_C \ket{m}_A\ket{n}_C = \sum_m c_m \ket{m}_A
\end{align}

Not surprisingly, the Choi state and its recovery of the channel is basis-specific, as the teleportation above makes explicit. However, its operational interpretation--that of state teleportation--we will find useful in applying shadow tomography concepts to process tomography.

\subsection{Creating a process shadow}

\begin{figure}[t]
    \centering
    \includegraphics[width=0.45\textwidth]{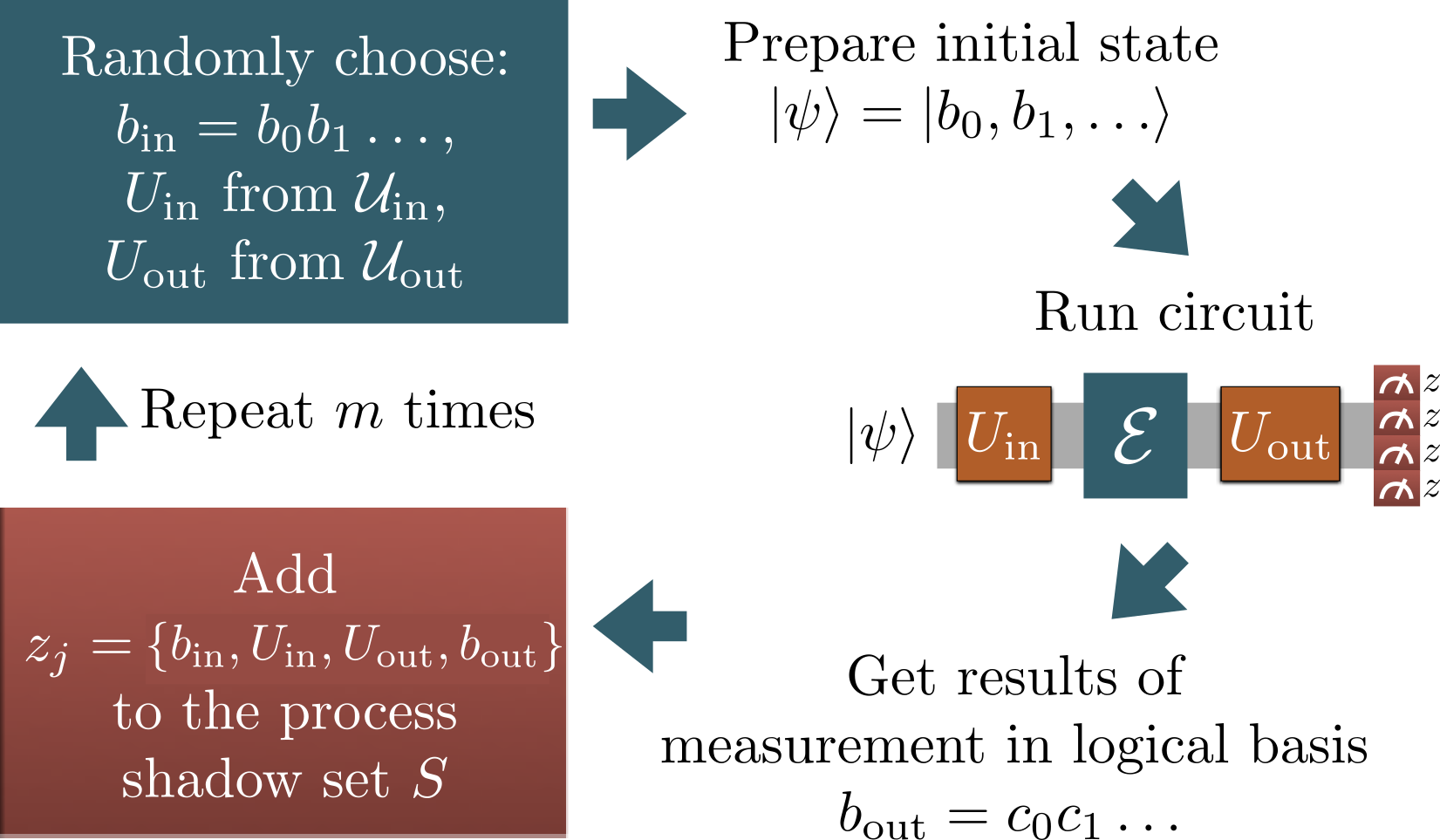}
    \caption{Our shadow process tomography algorithm. We prepare a random product state in the logical basis, apply a random unitary $U_\inp$ drawn from an ensemble $\Uin$, apply the channel $\mathcal E$, apply another random unitary $U_\out$ drawn from $\Uout$, and measure in the logical basis to obtain a bit string $\bout$. The combination $z = \{ \bin, U_\inp, U_\out, \bout\}$ forms a single-shot classical shadow of the channel. }
    \label{f:algorithm}
\end{figure}

The Choi isomorphism provides a mapping from quantum channels to density matrices. 
Therefore in theory, we can perform process shadow tomography by applying shadow tomography on the Choi state. 
However, such a procedure is practically inefficient because we would need to prepare the highly entangled state $\ket{\omega}$ between \emph{two} copies of the system. 
Instead, we use the teleportation interpretation of the Choi isomorphism and implement process shadow tomography using this procedure (see also \cref{f:algorithm}): 
\begin{enumerate}
    \item Uniformly draw a random bit string $\bin \in \{0,1\}^n$ and prepare $\ket{\bin}$
    \item Apply an unitary $U_\inp$ drawn randomly from an ensemble $\Uin$. In particular, we will consider two possibilities for $\Uin$: random single-qubit Pauli rotations ($\UP$) and random global Clifford ($\UC$).
    \item Apply the channel $\E$
    \item Apply an unitary $U_\out$ drawn randomly from an ensemble $\Uout$, which is not necessary the same as $\Uin$. Again, we will consider $\Uout = \UP$ and $\Uout = \UC$
    \item Measure in the computational basis to obtain a bit string $\bout$.
    \item Add the combination $z = \{ \bin, U_\inp, U_\out, \bout\}$ to the set $\zeta$ of shadow representations of the Choi state $\eta$ (\cref{eq:Choi-state}).
    \item Repeat the above $m$ times.
\end{enumerate}
Define 
\begin{align}
    \ket{z} = U_\inp^T \ket{\bin}&\otimes U_\out^\dag \ket{\bout}.\label{eq:z-def}
\end{align}
We show in \cref{sec:bin-prob} that the probability of getting a particular combination $z$ can be written as
\begin{align}
    P(z|\eta) = P(U_\inp,U_\out,\bin){\rm Tr}\big[ \ketbrad{z} \eta\big]
\end{align}
Then,  
\begin{align}
    \sum_z P(z|\eta)
    \ketbrad{z} = \mathcal M_{\Uin\otimes\Uout}(\eta)
\end{align}
is a linear map of the Choi state. 
If $\mathcal M_{\Uin\otimes\Uout}$ is invertible, we define
\begin{align}
    \hat \zeta = \mathcal M^{-1}_{\Uin\otimes\Uout}\left(\ketbrad{z}\right)
\end{align} 
as a single-shot shadow of $\eta$ labeled by its parameters $z = \{ \bin, \Uin, \Uout, \bout\}$. 
Clearly, taking the expectation over all possible such choices and outcomes for $z$ will yield the original Choi state; that is, $\mathbb E_{U_\inp,U_\out,\bin,\bout} \hat \zeta = \eta$.

Repeating the procedure above for $m$ times, we obtain a collection of $m$ shadows $\hat \zeta_1,\dots,\hat \zeta_m$. 
The average of the collection, $\zeta = \frac{1}{m} \sum_{j=1}^m \hat \zeta_j$, provides an estimator for the Choi state $\eta = \lim_{m\rightarrow \infty} \zeta$.

In shadow tomography, an important metric is the sample complexity, i.e. the minimum $m$ such that the estimator is close to correct values.
Requiring the distance between the state estimator $\zeta$ and the Choi state $\eta$ to be small is usually too restrictive (The sample complexity would scale polynomially with the Hilbert space dimension).
Instead, we only ask that the estimator and the Choi state are close in computing the expectation values of certain observables. 

Given a density matrix $\rho$ and an observable $O$, we define $\hat o  = \Tr[\hat \zeta  \rho^T \otimes O]$ as the single-shot estimator of $o = \Tr[\E(\rho)O]$.
Again, $\mathbb E\ \hat o = o$, where the mean is taken over all $z$ combinations ($\Uin,\Uout,\bin,$ and $\bout$).
Considering a collection of $m$ single-shot estimators $\hat o_1,\dots,\hat o_m$, we define $\tilde o(\rho,N,K)$ as the median-of-means estimator, computed by dividing the collection into $K$ subsets, each of size $N = m/K$, and taking the median of the means of the subsets. 


Before we prove the sample complexity of process shadow tomography, we would like to note its qualitative differences from state shadow tomography.
First, if we were to actually apply state shadow tomography on the Choi state, both $\bin$ and $\bout$ would have been measurement outcomes and their distributions depend on the Choi state $\eta$. 
In contrast, $\bin$ in our procedure is drawn from an ensemble independent of $\eta$. 
Effectively, our procedure performs state shadow tomography on the Choi state and post-selects on the first copy of the system being in $\ket{\bin}$. 
We show in \cref{sec:bin-prob} that this distinction disappears when we draw $\bin$ uniformly.

The second difference between state and process tomography is that the latter allows possibly different choices for $\Uin$ and $\Uout$. This enables four different mix-and-match scenarios when we restrict the ensembles to either $\UC$ or $\UP$.  
The following theorem, the central result of this work,  estimates the sample complexity in the four scenarios, providing an analog of \cref{thm:state-tomo} for process shadow tomography.



\begin{theorem}\label{thm:thm-sample-complexity}
Let $O_1,\dots,O_M$ be a collection of $M$ operators
and $\rho_1,\dots,\rho_L$ be a collection of $L$ other operators.
Given $\epsilon,\delta\in[0,1]$, let
\begin{align}
    &K = 2\log (2ML/\delta) , \\
    & N = \frac{34}{\epsilon^2} 4^n
    \max_{\substack{1\leq j\leq M \\ 1 \leq \ell \leq M} }
    f_{\Uin} (\rho_\ell) f_{\Uout} (O_j), \label{eq:samples-per-subset}
\end{align}
where  
\begin{align}
    &f_{\mathcal U} (O) \coloneqq \begin{cases}
        4^{\supp(O)}\norm{O}^2 & \text{ if $\mathcal U = \UP$},\\
        \norm{\mathcal S(O)} & \text{ if $\mathcal U = \UC$},
    \end{cases}\\
    &\mathcal S(O) \coloneqq 2\left\{[2\Tr(O)^2+ \Tr(O^2)] \mathbb I + 2\Tr(O)O + 2O^2 \right\},
\end{align}
$||.||$ denotes the operator norm (also called the spectral norm) \cite{huang_predicting_2020}, and $\supp(O)$ is the number of qubits $O$ acts nontrivially on.
Then, a collection of $m = NK$ shadows is sufficient to ensure that, with probability at least $1-\delta$, the median of means $\hat o_j(\rho_k,N,K)$ is a good estimate of $\Tr[\E(\rho_k)O_j]$, i.e.
\begin{align}\label{eq:est-error}
    \abs{\Tr[\E(\rho_k)O_j] - \hat o_j(\rho_k,N,K)} \leq \epsilon, 
\end{align}
for all $1\leq j\leq M$ and $1\leq k \leq L$.
\end{theorem}
Note that $\norm{\mathcal S(\rho)} \leq 14$ if $\rho$ is a normalized density matrix and $\norm{\mathcal S(O)}\leq 14 \ \rank(O)^2 \norm{O}^2$ in general.

\begin{proof}
Using the Choi isomorphism, we can view the collections of $M$ observables and $L$ density matrices as $ML$ ``observables'' of the form $2^n \,\!\rho_k^T \otimes O_j$ for $1\leq k\leq L$ and $1\leq j\leq M$. The leading factor of $2^n $ comes from the Choi state normalization, and will factor straightforwardly out of subsequent calculations to give the factor of $4^n$ in \cref{eq:samples-per-subset}. We need to prove \cref{thm:thm-sample-complexity} for 4 different cases corresponding to $\Uin,\Uout\in \{\UC,\UP\}$.

First, we consider $\Uin = \Uout = \UP$, which correspond to performing state shadow tomography on the Choi state using random Pauli measurements.
This case has been analyzed in Ref.~\cite{huang_predicting_2020}, which shows that 
\begin{align}
    N &= \frac{34}{\epsilon^2} \max_{\substack{1\leq \ell\leq L\\1\leq j\leq M }}   4^{\supp(\rho_\ell\otimes O_j)} \norm{2^n \rho_\ell^T\otimes {O_j}}^2\\
    &=\frac{34}{\epsilon^2} 4^n \max_{\substack{1\leq \ell\leq L\\1\leq j\leq M }}  4^{\supp(\rho_\ell\otimes O_j)}  \norm{\rho_\ell}^2\norm{O_j}^2
\end{align}
is sufficient for the statement of \cref{thm:thm-sample-complexity}.

To prove the other cases, we need the following lemma, which comes from the unitary 3-design property of $\UC$ (see \cref{sec:proof-lem-clifford-3-design-inverse} for a proof):
\begin{lemma}\label{lem:clifford-3-design-inverse}
    Given an operator $O$, we have
    \begin{align}
        \sum_{b\in\{0,1\}^{\otimes n}}
        \mathbb E_{U\sim \UC}
        U^\dag \ketbrad{b}U \bra{b}U\mathcal M_{\UC}^{-1}\left(O \right)U^\dag\ket{b}^2 \leq 2 \mathcal S(O).
    \end{align}
\end{lemma}
With \cref{lem:clifford-3-design-inverse}, we now prove \cref{thm:thm-sample-complexity} when $\Uin = \Uout = \UC$.
From \cref{thm:state-tomo}, \cref{thm:thm-sample-complexity} holds if
\begin{align}
    N &= \frac{34}{\epsilon^2} \norm{\rho^T\otimes O}_{\text{shadow},\UC\otimes \UC,\Omega_{\text{Choi}} }^2\\
    &= \frac{34}{\epsilon^2}\max_{\sigma\in \Omega_{\text{Choi}}} \sum_{b\in\{0,1\}^{2n}} \mathbb E_{U\sim\UC\otimes \UC} 
    \Tr\left[\sigma U^\dag \ketbrad{b}U \right]\nonumber\\
    &\qquad\qquad\qquad\times\bra{b}U(\mathcal M_{\UC}^{-1})^{\otimes 2} \left(2^n \rho^T\otimes O \right)U^\dag\ket{b}^2\\
    &\leq \frac{34}{\epsilon^2 }4^n \max_{\sigma\in \Omega_{\text{Choi}}} 
    \Tr\left[\sigma \mathcal S(\rho^T)\otimes \mathcal S(O)\right]\\
    &\leq \frac{34}{\epsilon^2}4^n \norm{\mathcal S(\rho^T)}\norm{\mathcal S(O)},
\end{align}
where $\Omega_{\text{Choi}}$ is the set of (normalized) Choi states and we have applied \cref{lem:clifford-3-design-inverse} twice.
Note that $\Tr(\sigma) = 1$.
Therefore, $N = \frac{34}{\epsilon^2}4^n \norm{\mathcal S(\rho^T)}\norm{\mathcal S(O)}$ is also sufficient for the statement of \cref{thm:thm-sample-complexity}.
Since $\norm{\mathcal S(\rho^T)} = \norm{\mathcal S(\rho)}$, \cref{thm:thm-sample-complexity} follows for  $\Uin = \Uout = \UC$.

Next, we consider the case $\Uin = \UC$ and $\Uout = \UP$.
Again, by \cref{thm:state-tomo}, \cref{thm:thm-sample-complexity} holds if 
\begin{align}
    &N = \frac{34}{\epsilon^2} \norm{2^n \rho^T\otimes O}_{\text{shadow},\UC\otimes \UP,\Omega_{\text{Choi}} }^2\\
    &\leq \frac{34}{\epsilon^2}4^n \max_{\sigma\in \Omega_{\text{Choi}}}\sum_{b_2}\mathbb E_{U_2\sim \UP} \Tr\left[\sigma \; (\mathcal S(\rho^T) \otimes U_2^\dag\ketbrad{b_2}U_2)\right]\nonumber\\
    &\qquad\qquad\qquad\qquad\times \bra{b_2}U_2 \mathcal M^{-1}_{\UP}(O)U_2^\dag\ket{b_2}^2,\label{eq:NUCUP}
\end{align}
where we have also used \cref{lem:clifford-3-design-inverse} to evaluate the average over $U_1\sim\UC$ and the sum over $b_1\in\{0,1\}^{n}$. 
Let 
\begin{align}
    \sigma_2 = \frac{1}{\norm{S(\rho^T)}} \Tr_1\left[\sigma \; (\mathcal S(\rho^T)\otimes  \mathbb I) \right].
\end{align}
It is straightforward to verify that $\sigma_2\geq 0$ and $\Tr(\sigma_2) \leq 1$ for all $\sigma\in\Omega_{\text{Choi}}$.
    Therefore, we can further upper bound the right-hand side of \cref{eq:NUCUP} by
    \begin{align}
        &\frac{34}{\epsilon^2}4^n \norm{\mathcal{S}(\rho^T)}\norm{O}_{\text{shadow},\UP}\nonumber\\
        &\leq \frac{34}{\epsilon^2} 4^n 4^{\supp(O)} \norm{\mathcal{S}(\rho^T)}\norm{O},
    \end{align}
    leading to \cref{thm:thm-sample-complexity}.
    
    The case $\Uin = \UP$ and $\Uout = \UC$ follows the exact same steps with the roles of $\rho^T$ and $O$ exchanged.
    This completes the proof of \cref{thm:thm-sample-complexity}.
\end{proof}

We have now established concrete formal bounds for shadow process tomography which are similar in form to those for shadow state tomography. What does this mean for estimating quantities of interest?
	
One obvious difference between Theorem 2 and the analogous state shadows result is the daunting factor of $4^n$ in \cref{eq:samples-per-subset}. Mathematically, this factor of $4^n$ follows from the fact that the Choi state of \cref{eq:Choi-state}, which takes normalized input states to normalized output states via \cref{eq:Choi-to-channel}, has trace $2^n$. In significant contrast to the state shadow case, this exponential factor affects even the use of semi-global Clifford measurements when $\rho, O$ in \cref{eq:est-error} are rank-1 operators, i.e. state projectors. To explain the reason for the difference, in the state shadows case, if we prepare a (physical, unit trace) Choi state in the lab, the expectation value of the projector $\rho^T\otimes O$ can be estimated from randomized measurement with a cost independent of the system size. This total overlap includes a contribution from the exponentially small overlap of the input register state with $\rho^T$. By contrast, the expectation value in \cref{eq:est-error} corresponds to the probability of finding the channel-evolved state $\mathcal{E}(\rho)$ to be in state $O$; since a physical quantum channel will act without any initial probability of accepting or rejecting the input state, the probability of interest in \cref{eq:est-error} is exponentially larger than its physical state analogue. 

Next, consider the choice for $\Uin$. 
From \cref{thm:thm-sample-complexity}, this choice will affect $f_{\Uin}(\rho)$.
If $\Uin = \UC$ and $\rho$ is a normalized density matrix, 
\begin{align}
    f_{\UC}(\rho) = \norm{\mathcal S(\rho)} \leq 14
\end{align}
is bounded for all $\rho$.
On the other hand, if $\Uin = \UP$, $f_{\UP}(\rho)$ grows exponentially with the support size of $\rho$.
So unless $\rho$ is supported on only a few sites, we should always choose $\Uin = \UC$ over $\UP$. Put in other terms, use of $\UP$ in the input space means that estimation of the effects of the channel work best on either high temperature states ($\rho$ has a polynomial expansion in local operators in the inverse temperature) or on mostly depolarized inputs (such as most qubits set randomly and only a few polarized). On the other hand, the effects of the channel on highly entangled or highly polarized input states are best determined by using $\UC$ on the input.

Next, for the choice of $\Uout$, we note that
\begin{align}
    f_{\UC}(O) \leq 14\ \rank(O)^2 \norm{O}^2.
\end{align}
Similarly to the state shadow tomography, we would choose
\begin{itemize}
    \item $\Uout = \UC$ if $O$ is a low-rank observable, e.g. the fidelity to a reference state, entanglement witness.
    \item $\Uout = \UP$ is $O$ is supported on a few qubits, e.g. local observables.
\end{itemize}

This becomes subtler and more important as we look to the next section, in which we consider a unique element of shadow process tomography: the ability to compose multiple channels together to create an estimate of the total channel thus developed. 

Finally, there is a fundamental difference between the Choi state as concept and the Choi state as actual entangled state sent through a channel. Specifically, the latter case allows all options from shadow tomography. The former case only allows measurements that are not entangled between inputs and outputs. We will not remark on this additional power at this time, other than to say that in quantum sensing and related protocols the additional benefit of having access to the entangled bits can be substantial in, e.g., multiple interaction or applications of the channel for estimation~\cite{pirandolaFundamentalLimitsQuantum2019}.

\section{Getting practical: Pauli preparation and measurements}

We now will consider some of the practical benefits and challenges of working with process shadows. We will focus on the case of the unitaries $U$ being draw from single qubit operations $\UP$, as this is the most experimentally relevant regime at present. In contrast to the explicit teleportation-based tomographic concept implicit in the Choi state, we instead start by randomly sampling from both $\UP$ and $b$ for the input strings. Let us denote the shadow process $\zeta$ as before. As may be obvious, adding to the list is straightforward by taking additional samples and then reweighting.
	
Given $\zeta$ and the corresponding set of $z$'s $S_m$, we can estimate the action of the channel on qubit-qubit correlation functions, for example. Using $S_m$ to create such estimates requires some care; borrowing from shadow state tomography, we will in general suggest use of the median-of-means approach for any local estimators, as detailed in Ref.~\cite{huang_predicting_2020}.

As this section specializes to the case of $\UP$, it is convenient to define
\begin{equation}
    \tau_{b,\mu} \equiv \mathcal{M}^{-1}_{\UPone}[\ketbrad{b_\mu}] = 3 \ketbrad{b_\mu} - \mathbb{I} 
\end{equation}
where $\mu = X, Y,$ or $Z$ and $b = \pm 1$. 

\subsection{Transition probabilities}

One possible application of the above would be to estimation of transition probabilities. For transitions between computational-basis eigenstates, say from $\ket{i} \to \ket{f}$, this can be done by estimating the expectation value of the projector $O = \ket{i}\!\bra{i} \otimes \ket{f}\!\bra{f}$ in the Choi state $\eta$. A simple computation gives
\begin{equation}
    \langle O \eta \rangle = \sum_a |\langle f | M_a | i \rangle|^2 = P(i \to f),
\end{equation}
where the $M_a$ form a Kraus representation for the channel $\mathcal{E}(\rho) = \sum_a M_a \rho M_a^{\dagger}$. If the initial and final states $\ket{i,f}$ are localized to a small number $k$ qubits and the rest of the system was considered in a high temperature (full mixed) state, then the projector $O$ is a localized operator, and the random Pauli measurements should efficiently predict these transition probabilities. On the other hand, for an initial vacuum plus few excitation state, random Clifford measurements would yield reasonable bounds on estimation.


\subsection{Multi-time correlation functions}
Ref \cite{huang_predicting_2020} notes that the Pauli measurement scheme is well suited for estimating spatial correlation functions, like the two point function $\langle \sigma^z_i \sigma^z_j \rangle = \Tr [\rho \sigma^z_i \sigma^z_j]$ for sites $i$ and $j$. Using shadows, we can extend this to calculation correlation functions in time between local operators: $\Tr[\mathcal{E}(\rho A) B]$. Note, however, that the shadow norm squared for this correlation function goes as $4^k$ where $k$ is the size of the support of $\rho A\otimes B $. As noted above, this adds a constraint that $\rho$ be composed mostly of low-weight strings. As a result it is useful to turn emphasis away from the initial state and to consider the case $\rho = \mathbb{I}$ and unitary $\mathcal{E}$. In this case we don't consider the correlation function with respect to a particular state, rather we look learning about how operators evolve in the Heisenberg picture. For a unitary $U$ corresponding to evolution for time $t$ we can estimate functions like
\begin{equation}
\Tr[\mathcal{E}(\rho A) B] \to \Tr[U A U^\dag  B] = \Tr[A B(t)]
\end{equation}
with a sample cost exponential in the support of $A \otimes B$. More details on this application (including to nonunitary channels) can be found in \cref{ap:multitime}.

\subsection{Unitarity verification}

Shadow tomography is capable of predicting functions which are both linear and non-linear in the quantum state. In particular, Ref.~\cite{huang_predicting_2020} gives an algorithm for estimating $\rm{tr}(O \rho^2)$ for an arbitrary operator $O$. This proceeds by finding a linear operator $\tilde{O}$ such that $\Tr (O \rho^2) = \Tr (\tilde{O} \rho \otimes \rho)$, where the first trace is on a Hilbert space $\mathcal{H}$ while the second is on $\mathcal{H} \otimes \mathcal{H}$. For example, we can estimate the purity of a state $\rho$ using $O = S$ the SWAP operator, because
\begin{equation}
    \Tr_{\mathcal{H}} \rho^2 = \sum_{ij} |\rho_{ij}|^2 = \Tr_{\mathcal{H} \otimes \mathcal{H}} [S \rho \otimes \rho].
\end{equation} 
Applied to the Choi state of a given channel, this can be used to verify if the channel is unitary, i.e. has a single Kraus operator.

To see this, note that a channel $\E$ is unitary if and only if its Choi state,
\begin{equation}
    \eta = \sum_{ij} \ket{i}\! \bra{j} \otimes \E(\ket{i}\! \bra{j})
\end{equation}
is maximally entangled. (To see it in one direction: if $\E$ is unitary, one easily checks that $\eta^2 = \eta$, up to normalization. The converse is slightly more tedious but straightforward). From the definition, any Choi state has the property that the partial trace over the output system $B$ gives the maximally mixed state for the input system $A$:
\begin{equation}
    \eta_A := \Tr_{B}\eta = \mathbb I_A.
\end{equation}
Therefore, if the total state $\eta$ is pure up to normalization, $\Tr \,\eta^2 = d^2$ (or equivalently $\eta^2 = d \eta$, where $d = {\rm dim}~H_A$), then we can conclude that the Choi state is maximally entangled, and thus that the channel $\epsilon$ is unitary.

Unfortunately, the SWAP operator $S$ is completely non-local: $k$ qubits partitioned into two sets of $k/2$ qubits, the SWAP operator acts on all $k$ qubits. Thus according to the estimates in \cite{huang_predicting_2020}, random Pauli measurements would require a number of samples of order $4^k$ to estimate $\Tr\, \eta^2$. This is the same requirement as doing full process tomography. Random Clifford measurements likewise require a sample size scaling exponentially with the number of qubits. It would be extremely interesting to determine if there exists a measurement set that can produce non-exponential scaling, or to prove that no such set can exist.

This task has two somewhat disparate applications in quantum gravity. One is to the black hole information paradox. The complexity of decoding Hawking radiation to determine the initial, pre-black hole state is believed to be exponential in the logarithm of the Hilbert space dimension \cite{2013JHEP...06..085H}. A more basic question is to simply check whether the formation and radiation process produces a unitary channel. In a very different regime, many recent proposals have been made to experimentally determine if Newtonian gravity can entangle meso-to-macroscopic objects in a lab \cite{2013arXiv1311.4558K,2017PhRvL.119x0401B,2017PhRvL.119x0402M,Carney:2018ofe,2021PRXQ....2c0330C}. Understanding whether the channel generated by the gravitational interaction is unitary or not is crucially important in determining the implications of these experiments \cite{Carney_2022}. These experiments typically involve just a few low-dimensional systems (e.g., two qubits \cite{2017PhRvL.119x0401B}), and so the above exponential scaling does not present a substantial difficulty.


\subsection{Application to an incoming shadow state}

What if we start with a shadow tomography estimator $\sigma$ to the input state, $\rho_{in}$, which was derived from sampling over $\UP$ using shadow state tomography? If we have the estimator for the Choi state $\zeta$ we need to calculate 
\begin{align}
    \sigma_{\rm out} {\rm Tr}[\sigma^T_{in} \zeta] &= \sum_{\mu,b,\mu',b',\nu,c} \prod_{ij} {\rm Tr}[\tau^{T}_{b_j,\mu_j} \tau_{b_j',\mu_j'}] \tau^{(i)}_{c_i,\nu_i}
\end{align}
where the sum is over the set $S_m$ and the similar set $R_k$ that defines $\sigma$.
Note that the transpose operation on $\tau$ simply flips the sign of the $Y$-portion of the Pauli representation, analogous to complex conjugation, for this simple channel example.

Fortunately, summing up over sample pairs is relatively simple, and the final state involves a sum over only the output samples $\nu, c$. This seems like we have a statement that shadows map to shadows. However, unlike our original shadows, this estimator for the output no longer has positive values in front of each $\tau$ product. Let us consider the quasiprobability term:
\begin{equation}\label{eq:weights}
    \frac{1}{2} {\rm Tr}[\tau^{T}_{b,\mu} \tau_{b',\mu'}] = \begin{cases}
        5/2, & \mu = \mu'\neq Y\ {\rm and}\ b=b' \\
        -2 , & \mu = \mu' \neq Y\ {\rm but}\ b\neq b'\\
        -2, & \mu = \mu' = Y\ {\rm and}\ b=b' \\
        5/2 , & \mu = \mu' = Y\ {\rm but}\ b\neq b' \\
        1/4, & {\rm otherwise}
    \end{cases}.
\end{equation}

This points to an interesting future research direction: better understanding the emergence of an (effective) sign problem in the application of a shadow to a shadow; we discuss this in more detail at the end of this section.

\subsection{Channel composition}

Just as we can apply a shadow process to a shadow state, consider what happens if we have process shadows for two channels, $\mathcal{X}$ and $\mathcal{Y}$, each with Choi state shadows $\eta_X, \eta_Y$. Can we construct an approximation to the channel $\mathcal{Y} \circ \mathcal{X}$?

Using the teleportation-based interpretation of the Choi state, it suffices to project onto a maximally entangled state between the output register of $X$, $B$, and the input register of $Y$, $A'$. Using the Choi representation of both processes, with set $S_m$ of $b,\mu,c,\nu$ values for the first channel and $S_{m'}$ of $b',\mu', c', \nu'$ values for the second channel, we find immediately
\begin{equation}\label{eq:ChannelComp}
    \sigma_{\rm con} = \sum_{(c,\nu,b,\mu),(b',\mu',c',\nu')} \prod_{ii'} \tau^{(i)}_{c,\nu} \tau^{(i')}_{c',\nu'} {\rm Tr}[ \tau_{b,\mu}^T \tau_{b',\mu'}] 
\end{equation}

\subsection{Distribution for random states}

Consider a random variable $s$ drawn uniformly from $S = \{ 5/2, 1/4, 1/4, 1/4, 1/4, -2\}$. We are interested in properties of 
\begin{equation}
    \hat \pi = \Pi_i^N s_i.
\end{equation}
This corresponds to composing two $N$-qubit channels with Pauli shadows according to Eqs. \cref{eq:weights}, \cref{eq:ChannelComp} without any assuming any correlations between qubits. While the signed weights in \cref{eq:weights} point to the existence of a sign problem, we want to investigate how important it is in practice to keep track of weights of small absolute value or of negative sign. First, we can ask what is the probability that $k$ of the $N$ samples are either $5/2$ or $-2$. We get
\begin{equation}
    p_{\rm big}(k) =  \binom{N}{k} (1/3)^k (2/3)^{N-k}.
\end{equation}

Then, given that we have $k$ drawn uniformly from $5/2$ and $-2$, we want to know how many $-2$'s we have, and if the overall number is even. The conditional averages are
\begin{align}
    p_{\rm odd}(k) &= \lim_{p\rightarrow 1/2} \sum_{j \in {\rm odd}} \binom{k}{j} (p)^j (1/2)^{k-j} \\
    &= \lim_{p\rightarrow 1/2} \frac{1}{2} \left[ (1/2 + p)^k - (1/2 - p)^k \right]= \frac{1}{2} \\
    p_{\rm even}(k) &= \sum_{j \in {\rm even}} \binom{k}{j} (1/2)^j (1/2)^{k-j} \\
    &= \lim_{p\rightarrow 1/2} \frac{1}{2} \left[ (1/2 + p)^k + (1/2 - p)^k \right] = \frac{1}{2}. 
\end{align}

From this we see that, for $k \geq 1$, we have an equal probability of having an even number of $-2$'s and an odd number of $-2$'s. We also have the average number of $-2$'s is independent of being even or odd and is $k/2$.

The total probability over all $S$ options for having a negative weight (an odd number of $-2$'s) is
\begin{equation}
    p_{\rm odd} = \frac{1}{2} \left( 1- (2/3)^N \right).
\end{equation}
What is the total probability of positive? We have the chance that no large values are chosen, $(2/3)^N$, and then we have the chance that some number are chosen, in which case we have a 1/2 chance of being positive, giving 
\begin{equation}
p_{\rm even}= 1/2 (1 + (2/3)^N),    
\end{equation}
as expected.

Thus we conclude that for random states, there is only a slight bias towards positive weights $\pi >0 $ as the number of qubits, $N$, becomes large. We can understand this bias as related to the probability of having no $|S|>1$ values at all, e.g., $(2/3)^N$. We cannot hope to avoid the sign problem by having only a few negative weights.

However, while the weights are only slightly biased towards being positive, the amplitude of positive states is more heavily biased. 
We rewrite $\pi = (-1)^l A$ where $ A>0$, we can look at $a = \log A$ as a random variable, which is correlated with $l \in \{ 0, 1 \}$. For $k>0$ large values ($|S|>1$) we have $l$ evenly distributed and $\bar{a}(k)$ corresponds to $(N-k) \log 1/4 + \frac{k}{2} \left( \log 5/2 + \log 2 \right)$. In the large $N$ limit, $a$ becomes normally distributed, which means that returning to the distribution of $|A|$, we have (except for a $(2/3)^N$ correction) a log normal distribution with log mean given by averaging $\bar{\bar{a}} = \sum_k \bar{a}(k) p_k \sim \frac{2N}{3} \log (1/4) + \frac{N}{6} \log 5$ and standard deviation $\sim \sqrt{N/6} \log 5$. The log normal distribution implies that a few, rare values will dominate the overall behavior of the weights. It is then worth investigating whether strategies based on importance sampling can accelerate the process of channel composition.






\section{Numerics}

\begin{figure}
    \centering
    \includegraphics[width=0.45\textwidth]{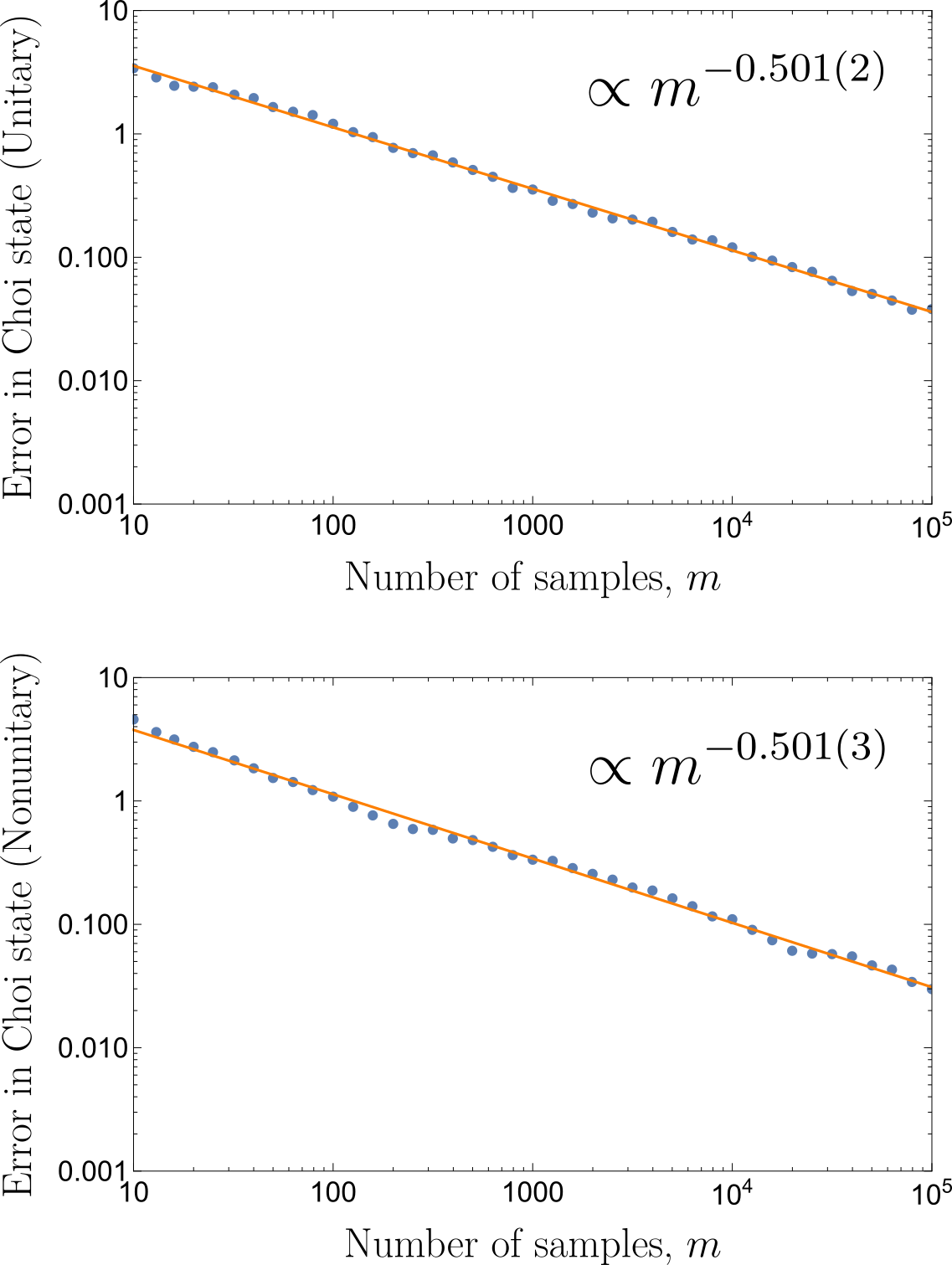}
    \caption{Numerical demonstration of the effectiveness of the experimental protocol for Pauli string operations on input and output for two qubits. Shadow representation of the Choi state from generating a Pauli bit string, acting on it with the channel, and measuring in random Pauli basis does converge to Choi state. For our error metric we use the operator norm, the absolute value of the largest eigenvalue of $\eta - \zeta$. Convergence goes as one over square root of sample number for both unitary (top) and general (bottom) channels. Data (blue points) and trendline (orange) are plotted on a log-log scale and correspond to shadows of a fixed channel. Insets in top right show average convergence exponent for ten such random channels.}
    \label{f:Numerics1}
\end{figure}
\begin{figure}
    \centering
    \includegraphics[width=0.45\textwidth]{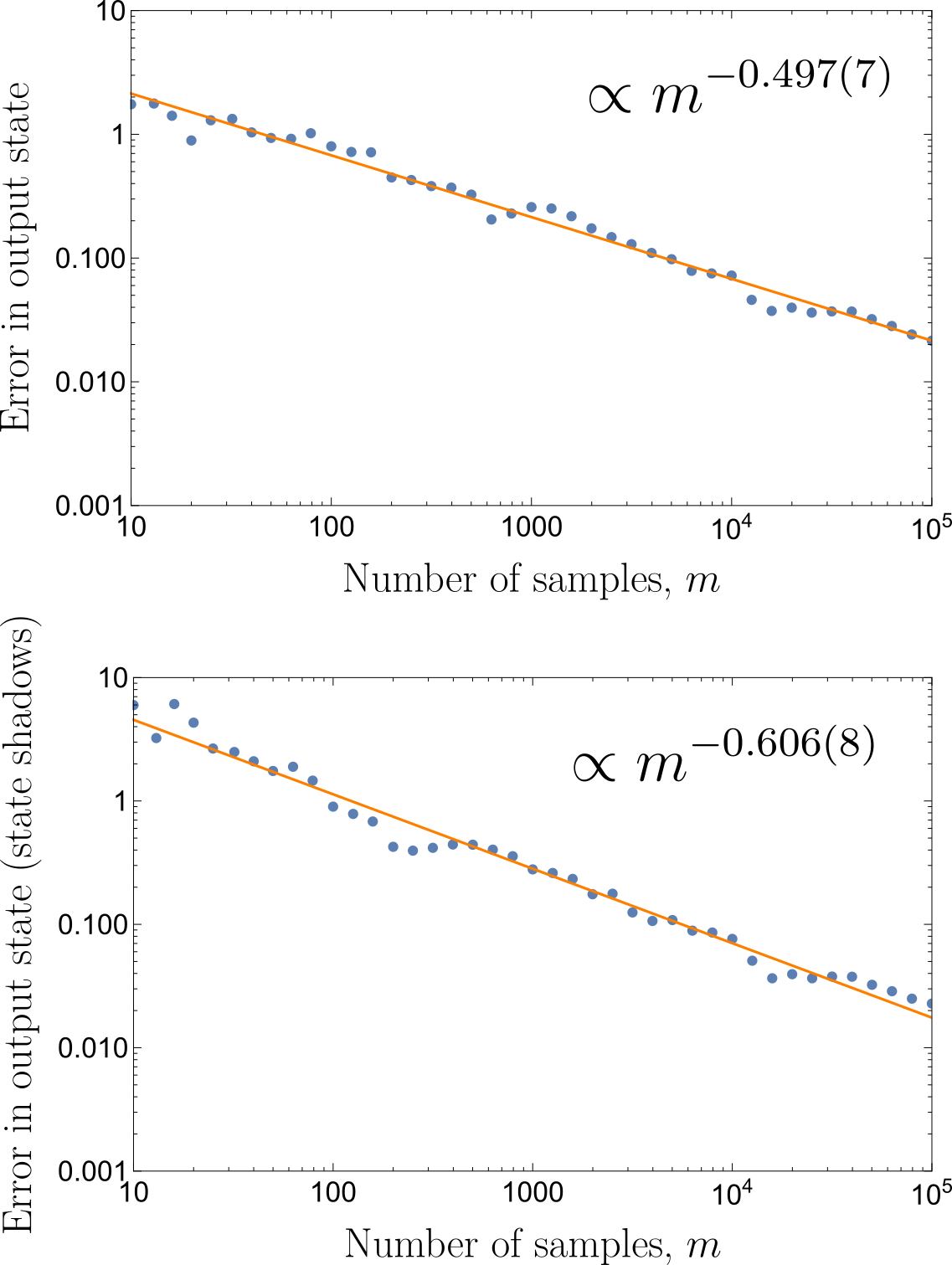}
    \caption{Convergence properties of shadow estimate of channel acting on a specific input density matrix for two qubits. We show convergence of our estimate of the output density matrix to the true density matrix using both the full input state (top) and a shadow representation of it (bottom). Error in calculating the output state is again the operator norm. Data (blue points) and trendline (orange) are plotted on a log-log scale and correspond to shadows of a fixed channel. Insets in top right show average convergence exponent for ten such random channels.}
    \label{f:Numerics2}
\end{figure}
\begin{figure}
    \centering
    \includegraphics[width=0.45\textwidth]{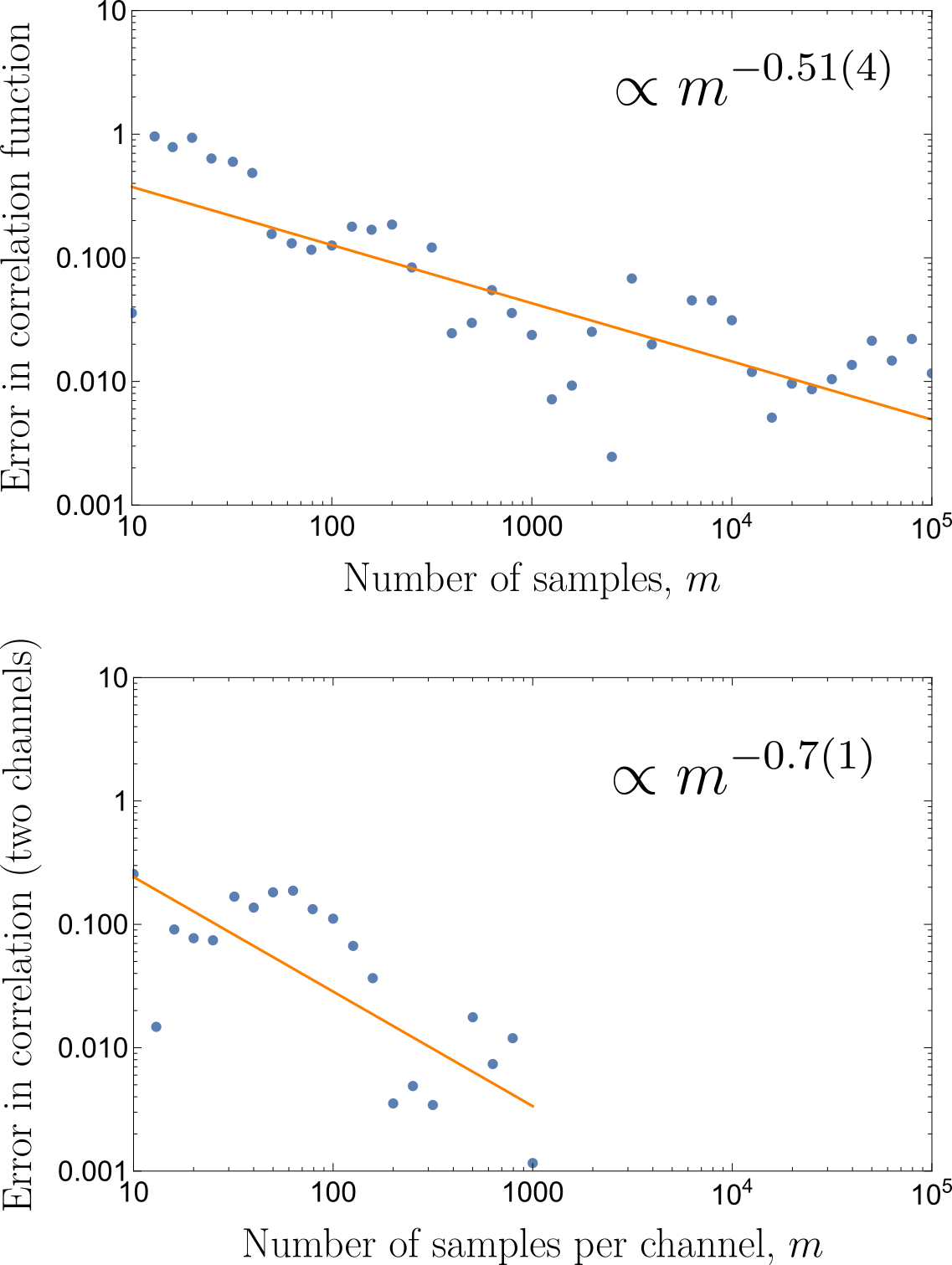}
    \caption{Estimating multitime correlation functions for two qubits. Error in the correlation function is the absolute value of the difference between the shadow calculation and the true value. We do this for a single channel (top) and for the composition of two channels (bottom). Data (blue points) and trendline (orange) are plotted on a log-log scale and correspond to shadows of a fixed channel or pair of channels. Insets in top right show average convergence exponent for ten such random channels.}
    \label{f:Numerics3}
\end{figure}

We now explore the capabilities of classical shadow process tomography with numerical simulations, providing conceptual and practical tests of the theory provided before. There are a few new things to investigate compared to previous numerical experiments in state tomography. First, as discussed above, there is the effect of a particular input state playing the role of an observable and thus contributing to the shadow norm. Second, there is the effect of channel composition.

In order to generate different Choi states so we can test the properties of a wide variety of channels (unitary, nonunitary, etc.), we make use of the numerical quantum information packages \texttt{qi} and \texttt{qinf050} \cite{mathematica_qi,mathematicaqinfo050}. The most significant use we make of these libraries is generating random unitary matrices which, using Stinespring dilation, is sufficient to generate arbitrary quantum channels. Below, in order to generate random nonunitary channels on $n_{sys}$ qubits, we generate a random unitary $U_{rand}$ on a system of $n_{sys}+n_{anc}$ qubits, where $n_{anc} = 2 n_{sys}$ is enough to give a full rank channel. We define the $j$th Kraus operator $K_j$ via
\begin{equation}
    K_j = \text{Tr}_{anc}[U_{rand} |0\rangle \langle j|\otimes \mathbb{I}_{sys}]
\end{equation}
with $|j\rangle$ the $j$th ancilla basis state. 

In this work, we work with systems of sufficiently small size that we can test convergence of the shadow representation of the Choi matrix to the full Choi matrix directly. As system size scales up, results from classical shadow state tomography suggest that shadow estimates of local correlation functions will converge much more rapidly than the actual state estimate. We also therefore investigate the convergence properties of these correlation functions. Below, when we speak of convergence of our estimate of an operator rather than of an expectation value, we always mean with respect to the operator norm, i.e. our error $\epsilon$ in some estimate $\zeta$ of an operator $\eta$ is $ \epsilon = ||\zeta - \eta||$.

In our first set of numerical experiments we investigate the properties of a single channel. In \cref{f:Numerics1}, we calculate the error between our shadow representation of a Choi state corresponding to a random channel. We see the expected convergence behavior, with no significant difference between taking shadows of Choi states corresponding to unitary vs. nonunitary channels. In \cref{f:Numerics2} we investigate the convergence properties of the output state resulting from tracing a specific input state against the Choi matrix. We generate a random density matrix with respect to the Hilbert-Schmidt measure, and subject it to a random channel. We then compare the convergence of the shadow estimate of the output operator to the actual operator evolved under the channel, first using the full input density matrix and second using a shadow estimator for the input density matrix. In the latter case, we fix the number of shadow samples of the input density matrix to be the same as the number of channel samples.

In \cref{f:Numerics3} we calculate the temporal correlation function
\begin{align}
    &\text{Tr}[\E( \rho_{in} \sigma_1^x ) \sigma_1^x] \label{Eq:correlationFunction}
\end{align}
where $\rho_{in}=|+\rangle \langle+|\otimes \frac{\mathbb{I}}{2}$, as described in Appendix C.
We use this input density matrix so we will have an operator made of low-weight Pauli strings, which is well suited to classical shadow estimation with local Pauli operations. We calculate the error by taking the absolute value of the difference between our correlation function calculated with shadows and the actual correlation function. 

We then calculate the same correlation function, but this time for a channel which is the composition of two channels for which we have shadow representations. Mathematically, one can concatenate channels in the Choi representation by projecting the output of the first channel and the input of the second into the maximally entangled state, as shown in \cref{f:choiiso}. We implement this procedure using the shadow representations of the two Choi states, using  \cref{eq:ChannelComp}. We use the same number of samples $m$ on each leg of the channel, so that when we concatenate, we do a sum over $m^2$ terms.
In the above numerics, we identify convergence behavior by assuming the error $\epsilon$ decays with the number of samples $m$ as a power law, i.e.
\begin{equation}
    \epsilon \propto m^{-b}
\end{equation}
We estimate the convergence exponent $b$ by doing linear regression on a log-log plot of the error versus number of samples, as the power law is determined by the slope of the line of fit. We find that errors in all cases with a single object represented by shadows scale as the number of samples raised to the -1/2 power, as expected from our main theorem. Curiously, combining state and channel shadows or channel and channel shadows, on the other hand, appear to converge faster. However we note that in the case of correlation functions with composed channels, the convergence exponent varies significantly across realizations, and it is not clear that a power-law fit really captures the convergence behavior.

\section{Outlook}
We have proposed a method for characterizing quantum processes using shadow process tomography. We have proved rigorous bounds on the performance of this method and conducted numerical experiments to analyze the ability of shadow tomography to predict specific features of a quantum channel. We have discussed why this technique suffers from an exponential cost in system size that is not present in the state tomography case. Some theoretical questions remain. For instance, what is the rigorous performance guarantee on applying shadows of process to shadows of quantum states? If the channel we seek to characterize has special properties (e.g. unitarity), can we use compressed sensing techniques to use the measurement record more efficiently? Can we use importance sampling to compose shadow representations of channels more efficiently? There are also opportunities in thinking about how to apply shadow process tomography in the near term. Perhaps using shadow process tomography to carry out channel composition in a computer, it is possible to propagate quantum simulation processes out past the experimental limits on runtime. Finally, these techniques may be a useful tool in verifying fidelity of complicated, multi-qubit gates.

\section{Acknowledgments}
We thank Victor Albert, Michael Gullans, and Srilekha Gandhari for helpful conversations, and Hsin-Yuan Huang and Zihao Li for pointing out an error in a previous versions of the manuscript. MCT acknowledges helpful discussions with Soonwon Choi, Wen Wei Ho, Di Luo,  Daniel Mark, and Tianci Zhou. MCT acknowledges Quantum Algorithms and Machine Learning grant from NTT, number AGMT DTD 9/24/20; JK and MCT were additionally supported by the Department of Energy, Office of Advanced Scientific Computing Research through the QOALAS program (grant 17-020469). DC is supported by the US Department of Energy under contract DE-AC02-05CH11231 and by the Quantum Information Science Enabled Discovery (QuantISED) for High Energy Physics grant KA2401032.

\textit{Note added:} In preparation of this manuscript we became aware of related work by Levy et al. \cite{levy2021classical}. Our results and those of \cite{levy2021classical} were obtained independently. We also note related work in \cite{helsen2021estimating}.

\section{Author Contributions}
JK and MCT contributed equally to this work.

\appendix

\section{Effects of the probability distribution of \texorpdfstring{$b_{\rm in}$}{bin}}
\label{sec:bin-prob}
In this section, we discuss the effect of the distribution of $\bin$ on the process shadow tomography.
As mentioned in the main text, if we were to perform the state shadow tomography on the Choi state $\eta$, the probability of obtaining a certain combination $z = \{\bin, \bout,U_\inp,U_\out\}$ would depend on $\eta$ as
\begin{align}
    P(z|\eta) = \frac{1}{2^n}P(U_\inp,U_\out)\Tr(\ketbrad{z} \eta), \label{eq:prob-bin-bout-state}
\end{align} 
where $\ket{z}$ is the state defined in \cref{eq:z-def} and the factor $1/2^n$ accounts for the renormalization of $\eta$.

In contrast to shadow state tomography where the probability of $\bin$ depends on $\eta$, we pick an $\eta$-independent probability distribution $P(\bin)$ for $\bin$ in our process shadow tomography.
Therefore, the probability of obtaining a combination $z$ is
\begin{align}
    &P(z|\eta) 
    = P(\bin,U_\inp,U_\out)P(\bout\vert \bin,U_\inp,U_\out,\eta)\nonumber\\
    &= 
    P(\bin,U_\inp,U_\out)  \frac{\Tr(\ketbrad{z}\eta)}
    {\Tr\left[\eta (U_\inp^T \ketbrad{\bin}U_\inp^*)\otimes \mathbb I \right]}\label{eq:prob-bin-bout-process},
\end{align}
where the denominator on the right-hand side is a normalization factor for the state resulted from projecting $U_\inp^* \otimes U_\out\,\eta\,(U_\inp^* \otimes U_\out)^\dag$ onto the subspace where the first copy of the system is in the state $\ket{\bin}$.

\Cref{eq:prob-bin-bout-state,eq:prob-bin-bout-process} are generally different.
However, because $\eta$ is a Choi state and the associated channel $\mathcal E$ is trace-preserving, the denominator on the right-hand side takes a trivial value:
\begin{align}
    \Tr\left[\eta (U_\inp^T \ketbrad{\bin}U_\inp^*)\otimes \mathbb I \right]
    = \Tr(\mathcal E(U_\inp \ketbrad{\bin}U_\inp^\dag)) = 1
\end{align}
for all $\bin$.
With $\bin$ being drawn uniformly at random ($P(\bin) = 1/2^n$), \cref{eq:prob-bin-bout-process} reduces to \cref{eq:prob-bin-bout-state} and we can treat $\bin$ as if it were the result of performing state shadow tomography on the Choi state.

A different way to show this is to consider measurement of the input qubits according to the Pauli operators set by the string $P_A = \prod_i \mu^{(i)}$ which operates only the input bits. For Choi states, they will always have the expectation value
\begin{align}
    \mean{P} =& {\rm Tr}[P_A \mathcal{I}_A \otimes \E_B \ketbrad{\omega}] \\
    =& \sum_{jmn} {\rm Tr}[ P_A \ket{m}\bra{n} ] {\rm Tr}[K_j \ket{m} \bra{n} K_j^\dag] \\
    =& \sum_{mn} {\rm Tr}[ P_A \ket{m}\bra{n} ] {\rm Tr}[ \sum_j K_j^\dag K_j \ket{m} \bra{n} ] \\
    =& \sum_m {\rm Tr}[P_A \ketbrad{m}] = {\rm Tr}[P_A] = 0,
\end{align}
where we write $\E_B(\rho) = \sum_j K_j \rho K_j^\dag$ using the Kraus representation. We also use the identity $\sum_j K_j^\dag K_j =\mathbb{I}$. Thus independent of the channel, there is no bias in the expectation value of any traceless observable on the input; by randomly choosing an input qubit state, we reproduce this unbiased input result.

\section{Proof of \texorpdfstring{\cref{lem:clifford-3-design-inverse}}{Lemma 1}}\label{sec:proof-lem-clifford-3-design-inverse}
We provide a proof of \cref{lem:clifford-3-design-inverse} in this section.
\begin{proof}
    Since $\UC$ is unitary 3-design, we have the following property~\cite[Lemma 7]{grossPartialDerandomizationPhaseLift2015}: 
    \begin{align}
        &\mathbb E_{U\sim \UC} U^\dag \ketbrad{b}U  = \frac{\mathbb I}{2^n},\label{eq:UC-1-design}\\
        &\mathbb E_{U\sim \UC} U^\dag \ketbrad{b}U \bra{b}U B U^\dag\ket{b}^2 \nonumber\\
        &= \frac{\Tr(B)^2 \mathbb I + 2\Tr(B)B + 2B^2 + \Tr(B^2)\mathbb I}{2^n(2^n+1)(2^n+2)},\label{eq:UC-3-design}
    \end{align}
    for any operator $B$.
    Using \cref{eq:inverse-clifford} for $\mathcal M_{\UC}^{-1}$ and the Cauchy-Schwarz inequality, we have
    \begin{align}
        &\sum_{b\in\{0,1\}^{ n}}
        \mathbb E_{U\sim \UC}
        U^\dag \ketbrad{b}U \bra{b}U\mathcal M_{\UC}^{-1} \left(O \right)U^\dag\ket{b}^2\nonumber\\
        &\leq 2 
        \sum_{b\in\{0,1\}^{\otimes n}}
        \mathbb E_{U\sim \UC}
        U^\dag \ketbrad{b}U \nonumber\\
        &\qquad\qquad\times\left[
        (2^n+1)^2\bra{b}U O U^\dag\ket{b}^2
        +\Tr(O)^2 
        \right]\\
        &\leq 2 \left[2\Tr(O)^2 \mathbb I + 2\Tr(O)O + 2O^2 + \Tr(O^2)\mathbb I\right].
    \end{align}
    Therefore, \cref{lem:clifford-3-design-inverse} follows.
\end{proof}


\section{Calculating multitime correlation functions with classical shadows}\label{ap:multitime}
We now examine how shadows can be used to estimate correlation functions. First, look at the correlator of the Pauli $x$ operator on site 1, $\sigma_1^x$,  and the time-delayed Pauli $y$ operator on site 2, $\sigma_2^y$, with unitary evolution for time $t$ by the operator $U$:
\begin{align}
    \langle \sigma^x_1(t_0) \sigma^y_2(t_0+t) \rangle &= \text{Tr}_{sys}[\rho_{sys}(t_0)\sigma^x_1 U^\dag \sigma^y_2 U] \nonumber \\
    &=\text{Tr}[U\rho_{sys}(t_0)\sigma^x_1 U^\dag \sigma^y_2 ].
\end{align}
This can be generalized to nonunitary channels \cite{carmichael1993},
\begin{align}
    \langle \sigma^x_1(t_0) \sigma^y_2(t_0+t) \rangle &=\text{Tr}_{sys}[\sum_i K_i\rho_{sys}(t_0)\sigma^x_1 K_i^\dag \sigma^y_2 ].
\end{align}
\indent
To motivate this generalization, we consider Stinespring dilation. We can represent a general completely positive trace-preserving map by unitary evolution $V$ acting on a combined system and bath, after which we trace out the bath
\begin{align}
    \rho_{sys} \to \sum_{i} K_i \rho_{sys}K_i^\dag = \text{Tr}_{Bath}[V \rho_{bath}\rho_{sys} V^\dag],
\end{align}
where we do not just get $\rho_{sys}$ back because the partial trace over operators on the composite space is in general not cyclic. With the addition of the bath, it is obvious that an analogous expression to Eq. 1 is the right one for the correlator
\begin{align}
    \langle \sigma^x_1(t_0) \sigma^y_2(t_0+t) \rangle&=\text{Tr}_{sys, bath}[\rho_{bath}\rho_{sys} \sigma_1^x V^\dag \sigma_2^y V] \nonumber\\
    &= \text{Tr}_{sys, bath}[V\rho_{bath}\rho_{sys} \sigma_1^x V^\dag \sigma_2^y ]\nonumber\\
    &=\text{Tr}_{sys}[\sum_i K_i\rho_{sys} \sigma_1^x K_i^\dag \sigma_2^y ].
\end{align}
We know how to apply the channel to $\rho_{sys}\sigma_x^1$ via the Choi isomorphism since it works for any matrix, not just density matrices. For any matrix $M$ we feed in the transpose
\begin{align}
    &\sum_i\text{Tr}_{anc}[\sum_{j_1,j_2} M_{j_2 j_1}|j_1\rangle\langle j_2 |\sum_{k,l} |k\rangle K_i|k \rangle\langle l|\langle l |K_i^\dag ] \nonumber\\
    &= \sum_{k,l} M_{kl}  K_i|k \rangle\langle l |K_i^\dag \nonumber\\
    &= K_i M K_i^\dag
\end{align}
and get out the channel applied to $M$.
\newline \indent 
\begin{widetext}
    Using process shadows to estimate this correlation function, the traces over the in register and out register factorize for each measurement result
    \begin{align}
        &\langle \sigma^x_1(0) \sigma_2^y(t)\rangle \nonumber\\
        &\approx\frac{d}{m}\sum_{n=1}^m \, \text{Tr}_{out}\Big[\;\text{Tr}_{in}\Big[\big(\rho_{in}\sigma_1^x\big)^\top \prod_{r\in \text{qubit labels}} \Big(3|b^{in,r}_{n}\rangle  \langle b^{in, r}_{\mu,n} | - \mathbb{I}\Big)\Big(3|b^{out,r}_{\mu,n}\rangle  \langle b^{out, r}_{\mu,n} | - \mathbb{I}\Big)\Big]\;\sigma^y_2\Big]\nonumber\\
        &=\frac{d}{m}\sum_n \, \text{Tr}_{in}\Big[\big(\rho_{in}\sigma_1^x\big)^\top \prod_{r\in \text{qubit labels}} \Big(3|b^{in,r}_{\mu,n}\rangle  \langle b^{in, r}_{\mu,n} | - \mathbb{I}\Big)\Big]\,\text{Tr}_{out}\Big[\prod_{r}\Big(3|b^{out,r}_{\mu,n}\rangle  \langle b^{out, r}_{\mu,n} | - \mathbb{I}\Big)\sigma^y_2\Big]\nonumber\\
        &=\frac{d}{m}\sum_n \text{Tr}_{in}\Big[\big(\rho_{in}\sigma_1^x\big)^\top \prod_{r\in \text{qubit labels}} \Big(3|b^{in,r}_{\mu,n}\rangle  \langle b^{in, r}_{\mu,n} | - \mathbb{I}\Big)\Big]3 b^{out, 2}_{\mu,n}\, \delta_{\mu,Y},
    \end{align}
\end{widetext}
and the trace over the output register drastically simplifies because we are time evolving a single site Pauli. Ultimately, this formula says (through the Kronecker delta $\delta_{\mu,Y}$) to throw out every output bitstring that did not measure site 2 along the $y$ axis, and if the measurement was along the $y$ axis to weight the corresponding input trace  with sign of the result.

\bibliography{apssamp}
	
\end{document}